# Orthorhombic boron oxide under pressure: *in situ* study by X-ray diffraction and Raman scattering

Kirill A. Cherednichenko,[1,2,3] Yann Le Godec,[2] Aleksandr Kalinko,[3,4] Mohamed Mezouar[5] and Vladimir L. Solozhenko [1,*]

[1] *LSPM–CNRS, Université Paris Nord, 93430 Villetaneuse, France*

[2] *IMPMC–CNRS, UPMC Sorbonne Universités, 75005 Paris, France*

[3] *Synchrotron SOLEIL, 91192 Gif-sur-Yvette, France*

[4] *Institute of Solid State Physics, University of Latvia, LV-1063 Riga, Latvia*

[5] *European Synchrotron Research Facility, 38043 Grenoble, France*

**Abstract**

*High-pressure phase of boron oxide, orthorhombic β-$B_2O_3$, has been studied in situ by synchrotron X-ray diffraction to 22 GPa and Raman scattering to 46 GPa at room temperature. The bulk modulus of β-$B_2O_3$ has been found to be 169(3) GPa that is in good agreement with our ab initio calculations. Raman and IR spectra of β-$B_2O_3$ have been measured at ambient pressure, all experimentally observed bands have been attributed to the theoretically calculated ones, and the mode assignment has been performed. Based on the data on Raman shift as a function of pressure, combined with equation-of-state data, the Grüneisen parameters of all experimentally observed Raman bands have been calculated. β-$B_2O_3$ enriched by $^{10}B$ isotope has been synthesized, and the effect of boron isotopic substitution on Raman spectra has been studied.*

## I. Introduction

The boron (III) oxide is known in vitreous form (g-$B_2O_3$) [1] and two crystalline forms: low-pressure α-$B_2O_3$ [1]-[3] and high-pressure β-$B_2O_3$ [3]-[5]. Vitreous boron oxide and α-$B_2O_3$ are important materials in ceramic and glass industries, mainly as components of special-purpose glasses [6]-[15]. The main interest in β-$B_2O_3$ is due to its unexpectedly high hardness (16±5 GPa [16]) comparable with that of the WC-10% Co hard metal [17]. In spite of the fact that the high-pressure phase of $B_2O_3$ has been rather intensively investigated [18]-[24], its properties still remain ill-understood, especially under pressure. For instance, the equation-of-state (EOS) measurements gave very different values of the β-$B_2O_3$ bulk modulus i.e. 90±15 GPa [22] and 169.9 GPa [24]. On the other hand, the phonon properties of β-$B_2O_3$ have not been studied so far, even at ambient pressure.

Here we report the results of *in situ* X-ray diffraction (XRD) and Raman studies of β-$B_2O_3$ under pressure to 46 GPa at room temperature. Based on *ab initio* calculations, we assigned the experimentally observed Raman and IR bands of β-$B_2O_3$ and estimated the Grüneisen parameters of all Raman-active phonon modes.

---

* Corresponding author: vladimir.solozhenko@univ-paris13.fr

## II. Experimental

Polycrystalline $\beta$-$B_2O_3$ has been synthesized from vitreous boron oxide, $g$-$B_2O_3$ according to the method described earlier [5]. $g$-$B_2O_3$ was produced by decomposition of $HBO_2$ metaboric acid (Aldrich, 99.0%) at 940 K in Ni crucible and subsequent re-melting (in order to remove air bubbles). $^{10}B$ enriched vitreous boron oxide was synthesized by thermal decomposition of $H_3^{10}BO_3$ boric acid (Cambridge Isotope Laboratories, Inc., 98%, $^{10}B$ enrichment 99%) to $H^{10}BO_2$, and then to $^{10}B_2O_3$. Orthorhombic $\beta$-$B_2O_3$ has been synthesized from remelted $g$-$B_2O_3$ in a toroid type high-pressure apparatus at 7.7 GPa and 1300 K with subsequent slow temperature decrease down to 700 K and quenching. Boron nitride (grade AX05, Saint-Gobain) capsules were used to isolate the reaction mixture from the graphite heater.

The X-ray diffraction studies (G3000 TEXT Inel diffractometer, Cu K$\alpha_1$ radiation) showed that all synthesized samples are highly crystalline $\beta$-$B_2O_3$ without any impurities (the lattice parameters are presented in Table I).

Raman spectra at ambient pressure have been measured in the 200-2500 cm$^{-1}$ range using Horiba Jobin Yvon HR800 Raman spectrometer. The 632.8 nm line of a He-Ne laser (10-µm beam spot) has been used for excitation; the laser power was less than 30 mW. A single crystal cubic Si has been used for spectrometer calibration.

The Fourier transform infrared absorption spectra in mid-infrared (450-4000 cm$^{-1}$) and far-infrared (10-600 cm$^{-1}$) regions were recorded using a Bruker IFS 125HR spectrometer. The $\beta$-$B_2O_3$ powders were mixed with KBr (for mid-infrared range) and polyethylene (for far-infrared range) and pressed in pellets.

*In situ* X-ray diffraction experiments up to 22 GPa were conducted at ID27 beamline (ESRF) in a Chervin-type membrane diamond anvil cell (MDAC) [25] equipped with 300-µm culet anvils. The polycrystalline powder sample was loaded into a 100-µm hole drilled in a rhenium gasket pre-indented down to ~25 µm. Neon was used as a pressure transmitting medium to maintain quasi-hydrostatic conditions. Pressure in a MDAC was determined from both the shift of the ruby fluorescence $R_1$ line [26],[27] and equation of state of neon [28],[29], the pressure drift at each pressure point did not exceed 0.4 GPa. The pressure values determined according to both pressure gauges (ruby and neon) are presented in Table II. Up to 22 GPa these values are very close indicating negligible strains and stresses and, consequently, insignificant pressure gradients all over the cell. High-brilliance focused (3 × 3 µm$^2$) synchrotron radiation was set to a wavelength of 0.3738(1) Å. Angle-dispersive X-ray diffraction patterns were collected using MAR 345 image plate detector with exposure time of 10 seconds. 2D diffraction patterns were processed with FIT2D [30] and GSAS [31] software. The lattice parameters and unit cell volume were refined using DICVOL04 and WinPLOTR packages in FullProf software [32],[33] and are presented in Table II.

High-pressure Raman studies were performed in a Le Toullec-type MDAC [34] (Re gasket, Ne pressure transmitting medium) using Ar$^+$ laser (514.5 nm, 5 µm beam spot). The spectra have been recorded in the 200-2000 cm$^{-1}$ range using Horiba Jobin Yvon HR460 Raman spectrometer. The spectrometer was calibrated using a single crystal cubic Si at ambient conditions. A laser power was 10 mW; no effect due to laser heating of the sample has been observed. Raman spectra of $\beta$-$B_2O_3$ have been measured in 17 points in the 0.3-46.4 GPa pressure range at room temperature. The sample pressure was determined from the shift of the ruby fluorescence $R_1$ line [27]; pressure drift during each experiment did not exceed 0.3 GPa.

## III. Calculations details

The first principles LCAO calculations for β-B$_2$O$_3$ have been performed using the CRYSTAL09 code [35]. The optimized in earlier calculations all-electron basis sets for boron [36] and oxygen [37] atoms have been used. In the CRYSTAL09 code, the accuracy of the calculation of the bielectronic Coulomb and exchange series is controlled by the set of tolerances, which were taken to be ($10^{-8}$, $10^{-8}$, $10^{-8}$, $10^{-8}$, $10^{-16}$). When the overlap between two atomic orbitals is smaller than $10^{-8}$, the corresponding integral is disregarded or evaluated in a less precise way [35]. The Monkhorst-Pack scheme [38] for an 8×8×8 k-point mesh in the Brillouin zone was applied. Self-consistent field calculations were performed for hybrid DFT–HF WCGGA–PBE-16% functional [39]. The percentage 16% defines the Hartree–Fock admixture in the exchange part of DFT functional.

We performed the full structure optimization procedure according to the energy minima criterion. The optimized structural parameters i.e. the lattice parameters (*a*, *b*, *c*) and unit cell volume ($V_0$) are listed in Table I.

The β-B$_2$O$_3$ equation-of-state parameters (bulk modulus $B_0$ and its first pressure derivative $B_0'$) have been estimated using routine implemented in the CRYSTAL09 code. To obtain the E(V) dependence, the unit cell volume was varied from 78.1% to 110.3% of the volume ($V_0$) corresponding to the energy minimum ($E_0$). The structure optimization was performed at each volume value. The calculated E(V) dependence have been fitted to the Murnaghan [40] (Eq. 1) and Birch-Murnaghan [41] (Eq. 2) EOSs with $V_0 = 149.23$ Å$^3$ and $E_0 = -29.933$ keV. The obtained $B_0$ and $B_0'$ values have been further used to estimate the P(V) dependence.

$$E(V) = E_0 + V_0 B_0 \left\{ \frac{1}{B_0'(B_0'-1)} \left(\frac{V}{V_0}\right)^{1-B_0'} + \frac{V}{V_0 B_0'} - \frac{1}{B_0'-1} \right\} \qquad (1)$$

$$E(V) = E_0 + \frac{9 V_0 B_0}{16} \left\{ \left[\left(\frac{V}{V_0}\right)^{\frac{2}{3}} - 1\right]^3 B_0' + \left[\left(\frac{V}{V_0}\right)^{\frac{2}{3}} - 1\right]^2 \left[6 - 4\left(\frac{V}{V_0}\right)^{\frac{2}{3}}\right] \right\} \qquad (2)$$

It should be noted, that electronic structure for reduced unit cell volume $V/V_0 < 0.78$ could not be calculated. According to Eqs. 1-4, the reduced unit cell volume of 0.78 corresponds to pressure of 58.2 GPa. Hence, the pressure dependencies of β-$^{11}$B$_2$O$_3$* unit cell parameters, $V_0$, etc. are given only up to 58.2 GPa.

The phonon frequencies for β-B$_2$O$_3$ containing 100% of $^{10}$B (β-$^{10}$B$_2$O$_3$*) and 100% of $^{11}$B (β-$^{11}$B$_2$O$_3$*) have been calculated using the direct (frozen-phonon) method [35],[42] at ambient pressure and selected pressure points up to 58.2 GPa using optimized geometries for corresponding reduced volume of unit cells.

## IV. Results and discussion

### A. *Equation of state of β-B$_2$O$_3$*

High-pressure phase of boron oxide has an orthorhombic unit cell (*Cmc2* space group) and consists of corner-linked distorted (BO$_4$)$^{5-}$ tetrahedral units, forming six- and eight-membered rings [5] (Fig. 1). The lattice parameters of β-B$_2$O$_3$ found in the present study are in good agreement with literature data [5]. The lattice parameters obtained from our *ab initio* calculations are also slightly larger than experimental ones. The deviations of the calculated *a*, *b*, *c* lattice parameters from the experimental ones (*β*-$B_2O_3$) are 0.35%, 0.50% and 0.17%, respectively. The unit cell volume deviation does not exceed 1%. It should be noted that there is no difference between lattice parameters of *β*-$^{10}B_2O_3$* and *β*-$^{11}B_2O_3$*, so further only *β*-$^{11}B_2O_3$* will be considered.

During compression of *β*-$B_2O_3$ a monotonous shift of the diffraction lines towards larger 2*θ*-values was observed (Fig. 2). The lattice parameters of *β*-$B_2O_3$ at different pressures are presented in Fig. 3. The most significant compression undergoes along *a* axis. We used the one-dimensional analog of the Murnaghan EOS (Eq. 3), to approximate the relation between the lattice parameters and pressure, in the same way as it has been done in [43]:

$$r = r_0 \left[1 + P\left(\frac{\beta'_{0,r}}{\beta_{0,r}}\right)\right]^{-\frac{1}{\beta'_{0,r}}} \tag{3}$$

Here *r* is the lattice parameter (index 0 refers to ambient pressure); $\beta_{0,r}$ is the axial modulus and $\beta'_{0,r}$ is the pressure derivative of $\beta_{0,r}$. Employing Eq. 4 and the axial moduli $\beta_{0,a}$, $\beta_{0,b}$ and $\beta_{0,c}$, the linear compressibilities ($k_r$) can be determined:

$$k_r = \beta_{0,r}^{-1} = \left(\frac{d \ln(r)}{d P}\right)_{P=0} \tag{4}$$

The $\beta_{0,r}$ and $k_r$ values from experimental and theoretical studies are presented in Table III. According to *ab initio* LCAO simulation (Fig. 3), only pressure dependence of *a* parameter perfectly fits with experimental data. The theoretical prediction for the compressions along *b* and *c* axes revealed the slights deviations from experimental points.

The EOS parameters of *β*-$^{11}B_2O_3$* have been estimated using the E(V) dependence obtained from LCAO calculations and approximated by Eqs. 1 and 2 (Fig. 4a). Volume dependence V(P) fitted to the Birch-Murnaghan EOS is shown in Fig. 4b. The isothermal bulk moduli ($B_0$) and their first pressure derivatives ($B_0'$) found from the approximations by Murnaghan and Birch-Murnaghan EOSs are 174.9±1 GPa, 2.1±0.1 and 173.1±1 GPa, 2.5±0.1, respectively. Both bulk modulus values are close to $B_0 = 169.9$ GPa reported in [24].

The experimentally observed unit cell volume change under compression is plotted in Fig. 5. For data fitting we used the Murnaghan (Eq. 5), third-order Birch-Murnaghan (Eq. 6) and Vinet [44] (Eq. 7) equations of state.

$$P(V) = \frac{B_0}{B_0'}\left[\left(\frac{V}{V_0}\right)^{B_0'} - 1\right] \tag{5}$$

$$P(V) = \frac{3B_0}{2}\left[\left(\frac{V_0}{V}\right)^{\frac{7}{3}} - \left(\frac{V_0}{V}\right)^{\frac{5}{3}}\right]\left\{1 + \frac{3}{4}(B_0' - 4)\left[\left(\frac{V_0}{V}\right)^{\frac{2}{3}} - 1\right]\right\} \tag{6}$$

$$P(V) = 3B_0 \frac{(1-X)}{X^2} e^{\left(1.5(B_0'-1)(1-X)\right)}$$

where $X = \sqrt[3]{\frac{V}{V_0}}$, $V_0$ is unit cell volume at ambient pressure $\tag{7}$

The corresponding $B_0$ and $B_0'$ values are presented in Table IV. $B_0$ values calculated using Vinet, Birch-Murnaghan, Murnaghan EOSs and our experimental data are very close to those reported in [24] and to our theoretical estimations. Such close $B_0$ values for these three independent datasets give strong reason to conclude that $B_0 = 90\pm15$ GPa reported in [22] is incorrect. The lower $\chi^2$ parameters for our dataset indicate the better accuracy of our data compared with data reported earlier [24].

In Fig. 5 the EOS-data for β-$B_2O_3$ obtained in the present (18 points) and previous [24] (13 points) studies are presented. One can see that these two independent datasets are complement to each other. Consequently, these two datasets have been combined into one which has been approximated by the same EOSs. The results of this fitting are listed in Table IV. In spite of the fact that the $\chi^2$ values for combined dataset are higher than those for separate datasets, they are still rather low. Approximation of the combined experimental P(V) data by the Murnaghan and Birch-Murnaghan EOSs gave the lowest $\chi^2$ values. Taking into account $\chi^2$ values presented in Table IV, the bulk modulus $B_0^t$ and its first pressure derivative $B_0^t{'}$ found for the combined P(V) dataset can be considered as rather precise and confident.

## B. Phonons study of β-$B_2O_3$ at ambient conditions

β-$B_2O_3$ has 10 atoms in the unit cell and according to group theoretical analysis, 30 phonon modes are expected at the Brillion zone center (Γ). They are described by the irreducible representation of the $C_{2v}$ point group:

$$\Gamma = 8A_1 + 7A_2 + 7B_1 + 8B_2 \tag{8}$$

Three modes among 30 are acoustic: $A_1 + B_1 + B_2$. 20 optical modes $7A_1 + 6B_1 + 7B_2$ are both infrared and Raman active, whereas $7A_2$ are only Raman active, thus there are 27 non-degenerate Raman active modes in Raman spectrum.

Raman spectra were measured in the 100-2500 $cm^{-1}$ frequency range, however all bands were observed in the 200-1200 $cm^{-1}$ range. For β-$B_2O_3$ only 12 bands were observed, whereas the Raman spectrum of β-$^{10}B_2O_3$ contains 6 bands (Fig. 6). The frequencies of all Raman bands of β-$B_2O_3$ and β-$^{10}B_2O_3$ listed in Table V perfectly match with those observed previously [21]. According to relative intensity, the Raman bands can be divided in three main groups: strong, medium and weak. For convenience we numbered all observed Raman bands i.e. #1, #2, #3, etc. The isotope substitution in β-$B_2O_3$ led to Raman bands shift toward high frequencies (Fig. 6) as well as it has been observed previously for r-BS [45]. In order to evaluate the isotope Raman shifts the following expression has been used:

$$\Delta\omega = \omega\ (β\text{-}^{10}B_2O_3) - \omega\ (β\text{-}B_2O_3) \tag{9}$$

The isotope shifts vary in the 0.3-15.8 $cm^{-1}$ range and are presented in Table V.

IR spectra of β-$B_2O_3$ (Fig. 7) have been measured for the first time. They were recorded in the 10-600 $cm^{-1}$ and 450-4000 $cm^{-1}$ frequency ranges and contain a vast quantity of bands. The intense bands in the 2400-4000 $cm^{-1}$ range can be explained by the presence of organic impurities (two bands at ~2850 $cm^{-1}$ and ~2920 $cm^{-1}$ referring to C-H stretching oscillations) and adsorbed water (broad band and shoulder at ~3210 $cm^{-1}$ and ~3380 $cm^{-1}$).

The results of our theoretical prediction of the phonon modes of β-$^{11}B_2O_3$* at T = 0 K are presented by the dashed lines in Figs. 6 and 7. The wavenumbers of all theoretically predicted phonon modes (Table V) are in good agreement with the experimentally observed bands. Based on the results of LCAO calculation 6 bands in far-infrared region and 10 bands in mid infrared region have been assigned to β-$B_2O_3$ (Table VI). Thus, an assignment of all experimentally observed Raman and IR bands of β-$B_2O_3$ to the phonon modes have been performed for the first time (Tables V and VI). The deviation of calculated Raman and IR modes from experimentally observed ones did not exceed 6%. The theoretically predicted isotope shifts (Δω*) vary in the 0.1-4.6 $cm^{-1}$ range (Table V). In spite of the fact that there is no good coincidence between absolute values of Δω and Δω*, the experimentally observed and theoretically predicted isotope shifts have the same order of magnitude. The Raman band at ~499 cm$^{-1}$ (marked by arrow in Fig. 6) is due to an impurity.

## C. Phonons study of β-B$_2$O$_3$ at high pressure

The Raman measurements of β-B$_2$O$_3$ have been performed at pressures to 46 GPa. Some of the observed Raman spectra are presented in Fig. 8. During compression all Raman bands expectedly shifted toward higher frequencies and revealed rather strong phonon's shift (Fig. 9). The *ab initio* prediction of the pressure dependencies of the phonon frequencies is in good agreement with experimental data. Three new bands appeared during compression: 608.6 cm$^{-1}$ at 11.3 GPa, 644.6 at 25.8 GPa and 701.4 cm$^{-1}$ at 26.9 GPa (further marked as #?). These bands were not observed in previous Raman studies of β-B$_2$O$_3$ [21]. Moreover, Raman bands (#1) and (#5) observed at 255.6 cm$^{-1}$ and 455.1 cm$^{-1}$ at ambient conditions disappear during compression at 11.3 and 37.0 GPa, respectively. Based on the literature data on β-B$_2$O$_3$ phase stability up to 41 GPa [18],[24], we excluded any possibility of phase transition. The bands appeared during compression might be assigned to B$_1$, B$_2$ and A$_2$ modes of β-B$_2$O$_3$ as it follows from our LCAO prediction of phonon modes under compression.

In order to express the extension of the phonon shifts observed experimentally and predicted theoretically toward high-frequency range, the quadratic equation has been employed for the approximation of experimental and theoretical data presented in Fig. 9:

$$\omega = \omega_0 + \omega_1 \cdot P + \omega_2 \cdot P^2 \qquad (10)$$

The parameters $\omega_1$ used in Eq. 10 are given in Table VII. It should be noted that the pressure dependency of Raman mode (#3) could be satisfactorily fitted only by the polynomial of the 3$^{rd}$ order, so $\omega_1$, $\omega_2$ and $\omega_3$ parameters are presented. The large variation of the $\omega_1$ values might be presumably explained by the fact that compression of the distorted BO$_4$ tetrahedron (the building block of β-B$_2$O$_3$) undergoes not uniformly. Although $\omega_1$ values found from experimental and theoretical pressure dependencies differ, the theoretical prediction of all phonon shifts under compression nicely matches with the experimental data. Based on a good agreement of pressure dependences for the experimental and theoretical data, the three unknown Raman bands (marked by #?) observed during compression have been attributed to B$_1$, B$_2$ and A$_2$ modes, respectively. Thus, the good agreement between theoretical and experimental pressure dependencies of the phonon modes has been used as a complimentary criterion for the mode assignment (Table IV).

In order to estimate the Grüneisen parameters of observed Raman modes under pressure we used the approximation of the experimental data presented in Fig. 9 by Eq. 11 as it has been done earlier [46]:

$$\omega = \omega_0 \cdot \left(1 + P \cdot \frac{\delta_0}{\delta'}\right)^{\delta'}$$

where $\delta_0 = \left(\frac{d \ln \omega}{dP}\right)_{P=0}$ and $\delta' = \delta_0^2 \left(\frac{d^2 \ln \omega}{dP^2}\right)_{P=0}$ \qquad (11)

A least-squares fit of Eq. 11 to the high-pressure experimental data yielded values of first-order parameters ($\delta_0$) for all observed Raman modes except the #?-modes appeared during compression. We defined the Grüneisen parameters using Eq. 12:

$$\gamma_G = B_0 \times \delta_0 \qquad (12)$$

where $B_0$ is a bulk modulus. We used the $B_0$ equal to 169.9 GPa (Vinet EOS) obtained from EOS measurements of $β$-$B_2O_3$. All $γ_G$ values are listed in Table V.

## V. Conclusions

300-K equation of state of orthorhombic boron (III) oxide, $β$-$B_2O_3$ has been *in situ* studied by angle-dispersive X-ray diffraction in DAC. The lattice parameters have been accurately determined at pressures to 22 GPa. The obtained V(P) dependence have been fitted to Vinet, Birch-Murnaghan and Murnaghan EOSs, and the corresponding values of bulk modulus ($B_0$) and its first derivative ($B_0′$) have been refined. The experimental data have been verified and proved by *ab initio* LCAO calculations.

$β$-$B_2O_3$ has also been studied by Raman spectroscopy at ambient and high pressure. *Ab initio* LCAO calculations have been performed in order to assign all experimentally observed phonons to the vibrational modes. The evolution of Raman bands of $β$-$B_2O_3$ has been studied up to 46 GPa at room temperature. Based on EOS data and pressure dependencies of the Raman modes, the corresponding Grüneisen parameters were calculated for the first time.

### Acknowledgements

The authors thank Dr. G. Le Marchand (IMPMC) for help in MDAC loading and V. Svitlyk (ESRF) for assistance in synchrotron X-ray diffraction experiments at beamline ID27 during beam time kindly provided by ESRF. This work was financially supported by the Agence Nationale de la Recherche (grant ANR-2011-BS08-018).

Table I. Cell parameters of $\beta$-B$_2$O$_3$, $\beta$-$^{10}$B$_2$O$_3$, $\beta$-$^{11}$B$_2$O$_3$* (LCAO, present work) and $\beta$-B$_2$O$_3$ [5].

| Parameter | $\beta$-B$_2$O$_3$ | $\beta$-$^{10}$B$_2$O$_3$ | $\beta$-$^{11}$B$_2$O$_3$* | $\beta$-B$_2$O$_3$ [5] |
|---|---|---|---|---|
| $a$, Å | 4.611(5) | 4.626(5) | 4.627(6) | 4.613(2) |
| $b$, Å | 7.804(5) | 7.824(3) | 7.843(8) | 7.803(4) |
| $c$, Å | 4.132(3) | 4.147(4) | 4.125(6) | 4.129(4) |
| $V_0$, Å$^3$ | 148.66 | 150.11 | 149.23 | 148.62 |

Table II. Experimental values of lattice parameters and unit cell volume of $\beta$-$B_2O_3$ as a function of pressure at room temperature.

| Pressure, GPa | | $a$, Å | $b$, Å | $c$, Å | Volume, Å$^3$ |
|---|---|---|---|---|---|
| Ne scale | Ruby scale | | | | |
| - | 0.0 | 4.612(4) | 7.804(7) | 4.132(5) | 148.66 |
| - | 0.3 | 4.607(1) | 7.798(1) | 4.129(1) | 148.34 |
| - | 0.5 | 4.606(1) | 7.796(1) | 4.128(1) | 148.21 |
| - | 1.2 | 4.599(1) | 7.787(1) | 4.122(1) | 147.63 |
| - | 2.1 | 4.589(1) | 7.776(1) | 4.117(1) | 146.91 |
| - | 3.2 | 4.577(1) | 7.762(2) | 4.112(1) | 146.06 |
| - | 4.5 | 4.561(1) | 7.741(1) | 4.101(1) | 144.81 |
| - | 5.7 | 4.548(1) | 7.724(2) | 4.097(2) | 143.92 |
| 7.2 | 7.3 | 4.532(1) | 7.706(3) | 4.088(2) | 142.77 |
| 8.7 | 8.9 | 4.514(2) | 7.697(3) | 4.081(1) | 141.78 |
| 10.5 | 10.5 | 4.496(1) | 7.670(1) | 4.072(1) | 140.43 |
| 11.3 | 11.4 | 4.486(1) | 7.656(1) | 4.067(1) | 139.68 |
| 12.7 | 12.8 | 4.473(1) | 7.640(1) | 4.059(1) | 138.73 |
| 14.6 | 14.6 | 4.455(1) | 7.615(2) | 4.048(2) | 137.32 |
| 16.5 | 16.8 | 4.431(1) | 7.589(3) | 4.036(3) | 135.71 |
| 18.8 | 18.6 | 4.418(2) | 7.572(3) | 4.029(2) | 134.78 |
| 20.3 | 20.2 | 4.398(5) | 7.563(10) | 4.025(5) | 133.89 |
| 22.3 | 21.6 | 4.394(2) | 7.541(4) | 4.022(3) | 133.25 |

Table III. Axial bulk moduli and linear compressibilities of $\beta$-$B_2O_3$ from experimental and theoretical studies.

| | Experiment | LCAO |
|---|---|---|
| $\beta_{0,a}$, GPa | 399±7 | 362±2 |
| $\beta_{0,b}$, GPa | 562±13 | 633±2 |
| $\beta_{0,c}$, GPa | 628±19 | 549±9 |
| $k_a \times 10^{-3}$ GPa$^{-1}$ | 2.51±0.05 | 2.76±0.01 |
| $k_b \times 10^{-3}$ GPa$^{-1}$ | 1.78±0.02 | 1.58±0.01 |
| $k_c \times 10^{-3}$ GPa$^{-1}$ | 1.59±0.05 | 1.82±0.03 |

Table IV. EOS parameters of $\beta$-$B_2O_3$ obtained from *ab initio* LCAO calculations and from the approximations of our experimental data, data of *Nieto-Sanz et al.* [24] and combined dataset using Vinet (*V*), Birch-Murnaghan (*B-M*) and Murnaghan (*M*) EOS. $\chi^2$ is an indication of the quality of the fit (lower for a better fit).

| EOS | Parameters | [24] | $\chi^2$ | Present study | $\chi^2$ | Combined dataset | $\chi^2$ | LCAO |
|---|---|---|---|---|---|---|---|---|
| V | $B_0$, GPa | 169.5±10 | 0.55 | 169.9±3 | 0.06 | 168.7±4 | 0.86 | - |
|   | $B_0'$ | 2.6±0.7 |   | 2.4±0.4 |   | 2.6±0.3 |   | - |
| B-M | $B_0$, GPa | 167.4±8 | 0.41 | 169.4±3 | 0.07 | 167.5±3 | 0.48 | 173.1±1 |
|   | $B_0'$ | 2.8±0.5 |   | 2.6±0.3 |   | 2.8±0.2 |   | 2.5±0.1 |
| M | $B_0$, GPa | 172.1±8 | 0.39 | 170.6±3 | 0.06 | 170.6±3 | 0.48 | 174.9±1 |
|   | $B_0'$ | 2.3±0.6 |   | 2.3±0.3 |   | 2.3±0.3 |   | 2.1±0.1 |

Table V. The frequencies of Raman bands ($\omega_0$) of $\beta$-$B_2O_3$ and $\beta$-$^{10}B_2O_3$ experimentally observed in the present study, and $\beta$-$^{11}B_2O_3$* and $\beta$-$^{10}B_2O_3$* phonon frequencies ($\omega_t$) theoretically predicted by LCAO calculations. The superscripts to the experimental wavenumbers (w, m, s) indicate the relative intensity of the Raman bands. The theoretically predicted ($\Delta\omega$*) and experimentally observed ($\Delta\omega$) isotope shifts of Raman bands are also presented.

| No. | Modes | Wavenumber (cm$^{-1}$) | | | | | | $\gamma_G$ |
|---|---|---|---|---|---|---|---|---|
| | | Experiment | | | LCAO | | | |
| | | $\omega_0$ ($\beta$-$B_2O_3$) | $\omega_0$ ($\beta$-$^{10}B_2O_3$) | $\Delta\omega$ | $\omega_t$ ($\beta$-$^{11}B_2O_3$*) | $\omega_t$ ($\beta$-$^{10}B_2O_3$*) | $\Delta\omega$* | |
| (#1) | A$_1$ | 255.6$^m$ | - | - | 262.3 | 263.1 | 0.8 | - |
| (#2) | B$_1$ | 288.6$^s$ | 289.3$^s$ | 0.7 | 288.9 | 289.1 | 0.2 | 3.993 |
| (#3) | A$_1$ | 333.5$^m$ | 334.1$^s$ | 0.6 | 343.0 | 344.6 | 1.6 | 1.430 |
| (#4) | A$_2$ | 385.7$^m$ | - | - | 402.6 | 406.9 | 4.3 | 0.593 |
| (#5) | B$_1$ | 454.6$^w$ | - | - | 436.1 | 436.6 | 0.5 | 1.426 |
| (#6) | A$_1$ | 525.3$^m$ | 527.5$^s$ | 2.2 | 530.8 | 534.5 | 3.7 | 0.518 |
| (#7) | A$_2$ | 555.7$^w$ | - | - | 568.7 | 578.8 | 10.1 | 0.635 |
| | B$_1$ | - | - | - | 604.5 | 609.6 | 5.1 | - |
| | B$_2$ | - | - | - | 608.6 | 613.4 | 4.8 | - |
| | A$_2$ | - | - | - | 699.9 | 717.8 | 17.9 | - |
| (#8) | A$_1$ | 706.2$^w$ | 722.0$^w$ | 15.8 | 714.8 | 733.3 | 18.5 | 1.672 |
| | B$_2$ | - | - | - | 733.9 | 756.2 | 22.3 | - |
| (#9) | A$_1$ | 786.0$^m$ | 786.3$^w$ | 0.3 | 800.3 | 813.0 | 12.7 | 1.862 |
| | B$_2$ | - | - | - | 831.9 | 851.5 | 19.6 | - |
| (#10) | A$_2$ | 816.7$^w$ | - | - | 846.5 | 867.2 | 20.7 | 0.674 |
| | B$_1$ | - | - | - | 873.0 | 896.3 | 23.3 | - |
| (#11) | A$_2$ | 879.5$^s$ | 885.9$^m$ | 6.4 | 889.4 | 910.1 | 20.7 | 0.615 |
| | A$_1$ | - | - | - | 904.6 | 922.4 | 17.8 | - |
| | B$_2$ | - | - | - | 904.7 | 930.3 | 25.6 | - |
| (#12) | B$_1$ | 952.7$^w$ | - | - | 936.1 | 941.1 | 5 | 0.762 |
| | A$_2$ | - | - | - | 971.6 | 997.9 | 26.3 | - |
| | B$_2$ | - | - | - | 1032.1 | 1064.9 | 32.8 | - |
| | B$_1$ | - | - | - | 1087.5 | 1136.1 | 48.6 | - |
| | B$_1$ | - | - | - | 1129.4 | 1180.3 | 50.9 | - |
| | A$_1$ | - | - | - | 1214.7 | 1271.6 | 56.9 | - |
| | B$_2$ | - | - | - | 1308.3 | 1331.9 | 23.6 | - |
| | A$_2$ | - | - | - | 1309.4 | 1332.2 | 22.8 | - |

Table VI. The phonon frequencies of $\beta$-$B_2O_3$ experimentally observed by IR spectroscopy ($\omega_0$) and theoretically predicted by LCAO calculations ($\omega_t$). Only IR-active modes are presented.

| No. | Mode | Wavenumber (cm$^{-1}$) | | No. | Mode | Wavenumber (cm$^{-1}$) | |
| --- | --- | --- | --- | --- | --- | --- | --- |
| | | Experiment | LCAO | | | Experiment | LCAO |
| | | $\omega_0$ ($\beta$-$B_2O_3$) | $\omega_t$ ($\beta$-$^{11}B_2O_3$*) | | | $\omega_0$ ($\beta$-$B_2O_3$) | $\omega_t$ ($\beta$-$^{11}B_2O_3$*) |
| | $A_1$ | | 262.3 | (#10) | $B_2$ | 819.9 | 831.9 |
| (#1) | $B_1$ | 297.6 | 288.9 | | $B_1$ | | 873.0 |
| (#2) | $A_1$ | 333.9 | 343.0 | (#11) | $A_1$ | 901.9 | 904.6 |
| (#3) | $B_1$ | 454.5 | 436.1 | | $B_2$ | | 904.7 |
| (#4) | $A_1$ | 527.4 | 530.8 | (#12) | $B_1$ | 939.2 | 936.1 |
| (#5) | $B_1$ | 588.2 | 604.5 | (#13) | $B_2$ | 1017.6 | 1032.1 |
| (#6) | $B_2$ | 603.9 | 608.6 | (#14) | $B_1$ | 1059.1 | 1087.5 |
| (#7) | $A_1$ | 704.2 | 714.8 | (#15) | $B_1$ | 1119.5 | 1129.4 |
| (#8) | $B_2$ | 714.8 | 733.9 | (#16) | $A_1$ | 1193.7 | 1214.7 |
| (#9) | $A_1$ | 783.9 | 800.3 | | $B_2$ | | 1308.3 |

Table VII. The parameters of Eq. 10 describing pressure dependencies of phonon shifts.

| No. | Mode | Experimental data | | | LCAO data | |
|---|---|---|---|---|---|---|
| | | $\omega_0$, cm$^{-1}$ | $\omega_1$, cm$^{-1}$GPa$^{-1}$ | $\omega_2$, cm$^{-1}$GPa$^{-1}$ | $\omega_t$, cm$^{-1}$ | $\omega_1$, cm$^{-1}$GPa$^{-1}$ |
| (#1) | $A_1$ | 255.6 | - | - | 262.3 | 3.21 |
| (#2) | $B_1$ | 288.6 | 2.69 | - | 288.9 | 3.92 |
| (#3) | $A_1$ | 333.5 | -1.82 | 0.40 | 343.0 | 1.01 |
| (#4) | $A_2$ | 385.7 | -1.00 | - | 402.6 | -1.78 |
| (#5) | $B_1$ | 454.6 | 3.62 | - | 436.1 | 3.90 |
| (#6) | $A_1$ | 525.3 | 1.55 | - | 530.8 | 1.04 |
| (#7) | $A_2$ | 555.7 | 1.94 | - | 568.7 | 1.88 |
| (#?) | - | 608.6[1] | 1.69 | - | - | 1.94 |
| (#?) | - | 644.6[2] | 3.25 | - | - | 1.12 |
| (#?) | - | 701.4[3] | 0.90 | - | - | 1.08 |
| (#8) | $A_1$ | 706.2 | 4.70 | - | 714.8 | 2.84 |
| (#9) | $A_1$ | 786.0 | 5.30 | - | 800.3 | 4.30 |
| (#10) | $A_2$ | 816.7 | 3.00 | - | 846.5 | 3.65 |
| (#11) | $A_2$ | 879.5 | 3.19 | - | 889.4 | 4.33 |
| (#12) | $B_1$ | 952.7 | 3.90 | - | 936.1 | 3.06 |

---

[1] at 11.3 GPa
[2] at 25.8 GPa
[3] at 26.9 GPa

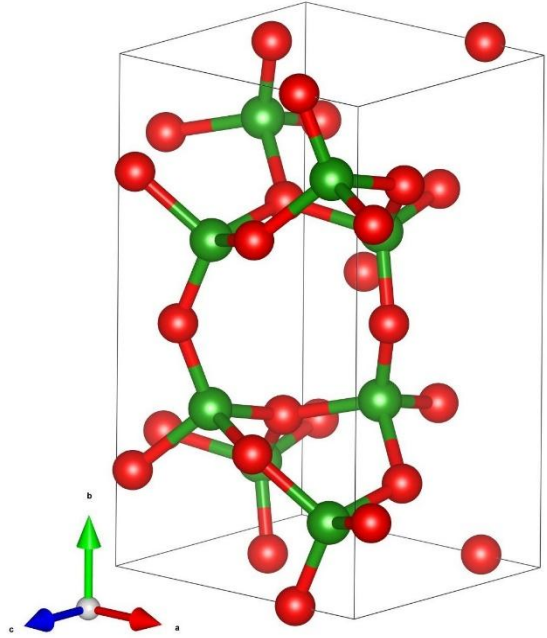

FIG. 1. Orthorhombic unit cell of $\beta$-$B_2O_3$ [5] (boron atoms are green, oxygen atoms are red).

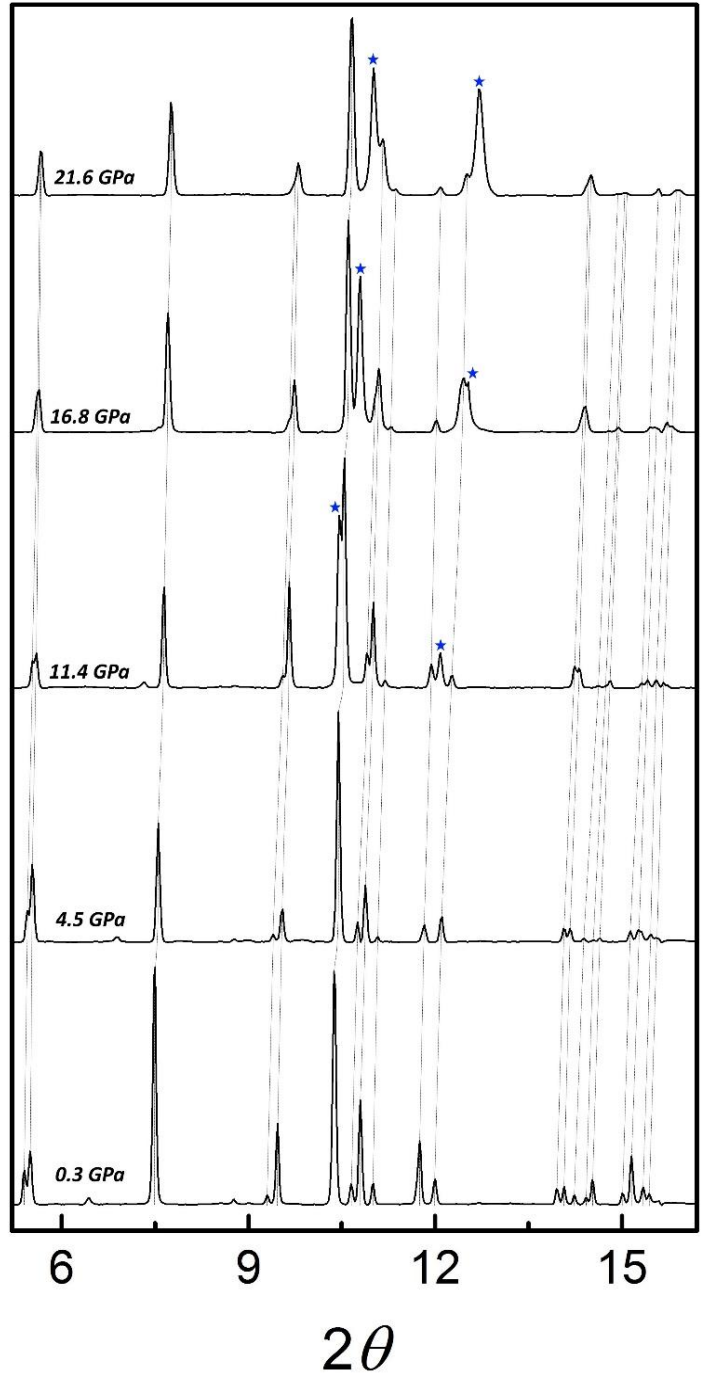

FIG. 2. Synchrotron X-ray diffraction patterns of *β*-B$_2$O$_3$ taken at different pressures (λ = 0.3738 Å). The shifts of *β*-B$_2$O$_3$ peaks under compression are traced by the dashed lines; Ne peaks are marked by blue stars.

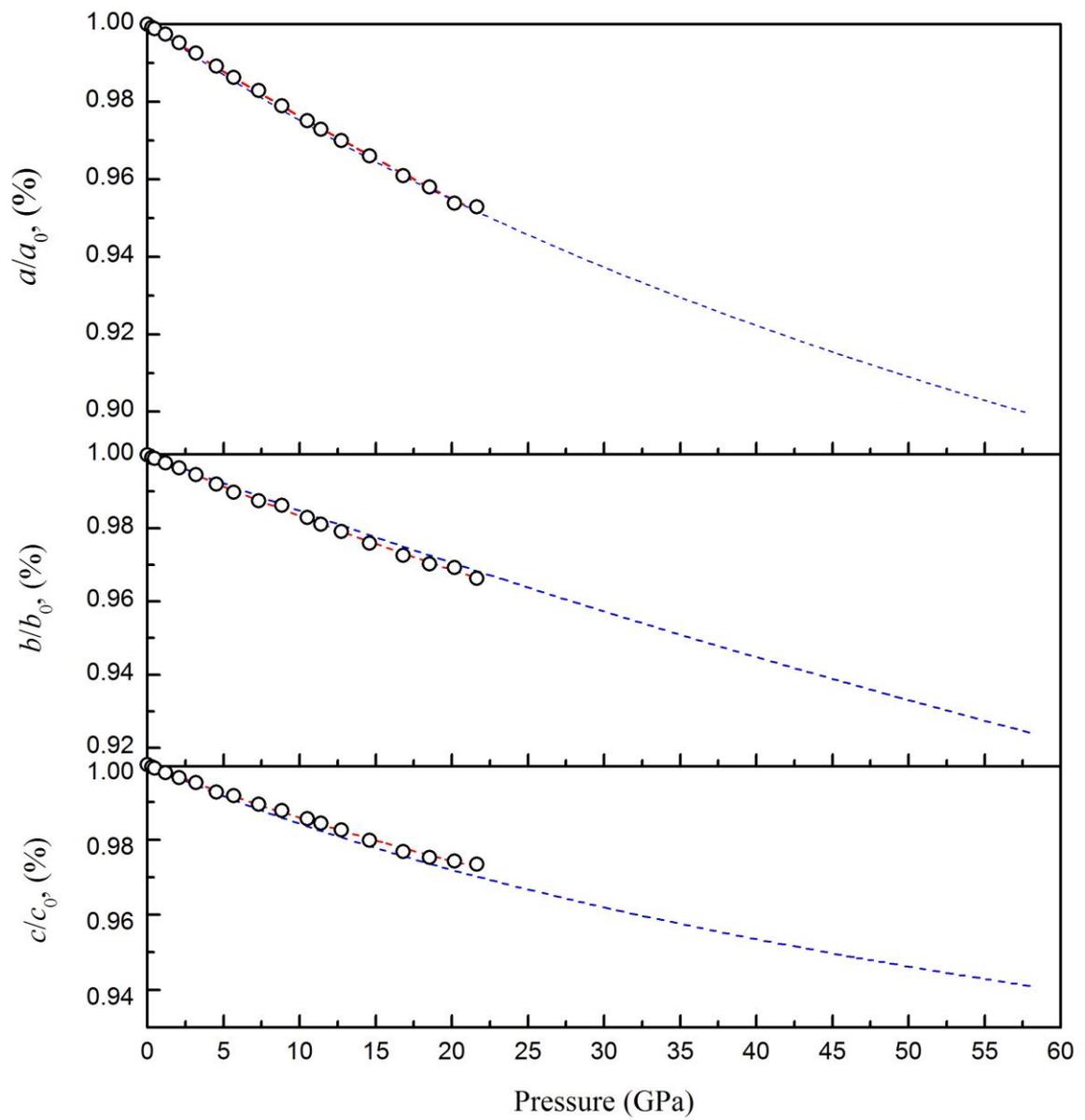

FIG. 3. Relative lattice parameters of *β*-$B_2O_3$ *versus* pressure. Open black circles represent experimental data. The red and blue dashed lines represent the fits of one-dimensional analog of the Murnaghan EOS to the experimental and LCAO data; the pressure values are given by the ruby gauge.

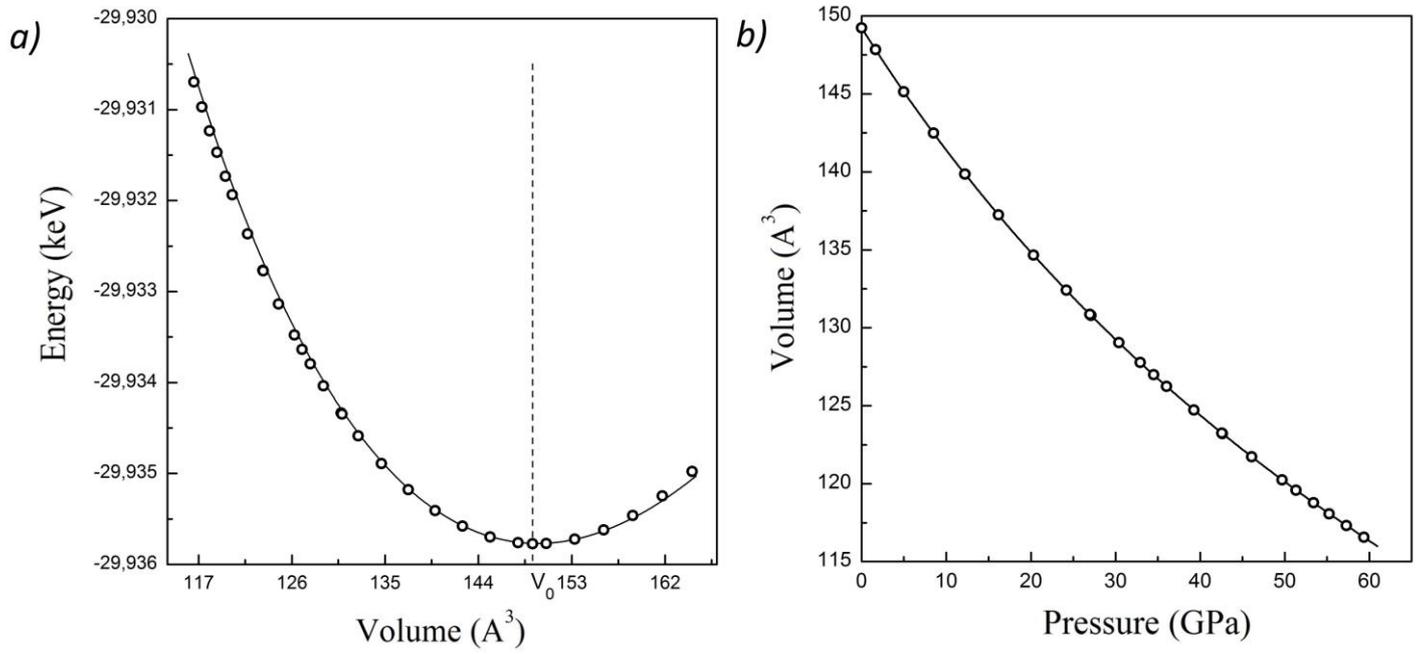

FIG. 4. (*a*) Energy variation of *β*-$B_2O_3$ unit cell *versus* volume (solid line is the Birch-Murnaghan fit, the $V_0$ is indicated by dashed line);  (*b*) *β*-$^{11}B_2O_3$* EOS, solid line is the Birch-Murnaghan fit to the theoretical data.

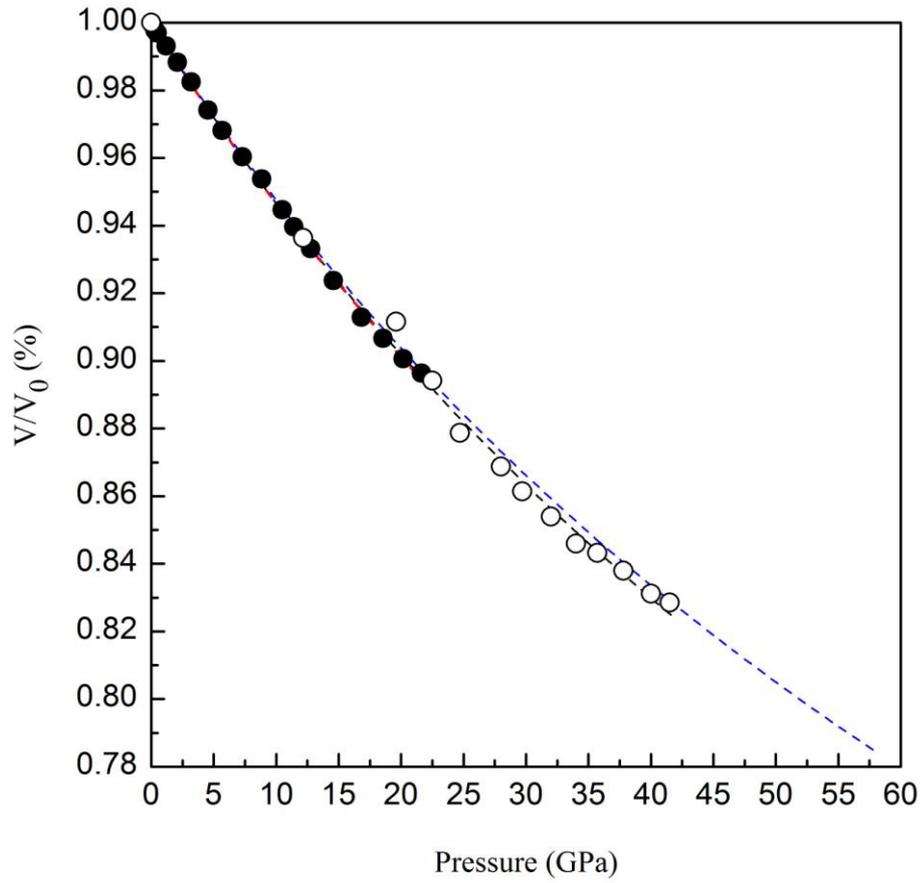

FIG. 5. Equation of state of β-$B_2O_3$. The solid and open circles represent our and previous [24] experimental data, respectively. The red solid and black dashed lines present Vinet fit to our experimental dataset and dataset [24], respectively; the blue dashed line presents the Birch-Murnaghan fit to the LCAO data; the pressure values are given by the ruby gauge.

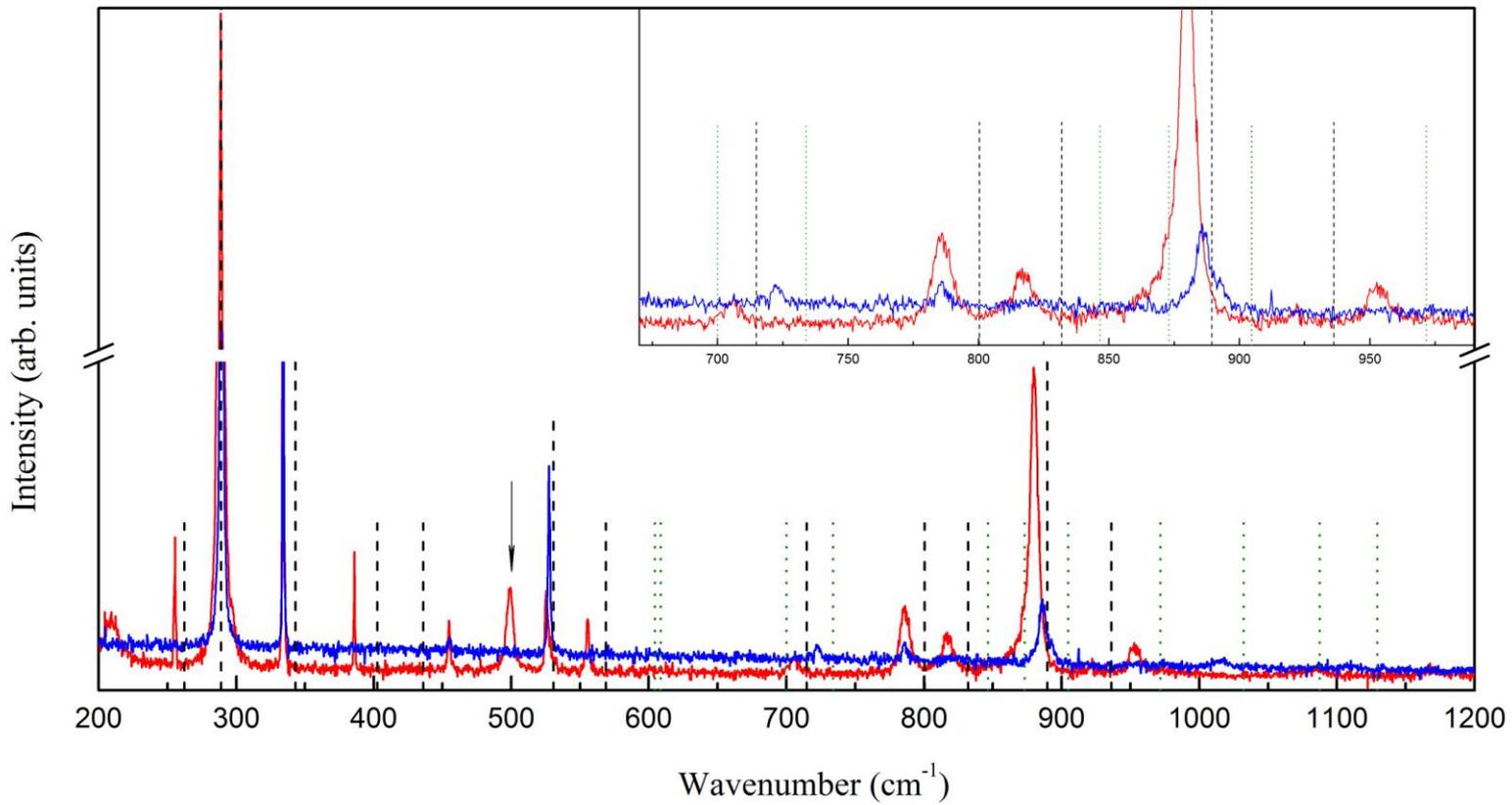

FIG. 6. Raman bands of $\beta$-$B_2O_3$ (red) and $\beta$-$^{10}B_2O_3$ (blue) at ambient conditions. The positions of the phonons predicted by LCAO calculations are marked by the lines: the black dashed lines correspond to the experimentally observed phonons, the green dot lines correspond to the non-experimentally observed phonons. The arrow indicates the Raman band related to a contamination. Inset: magnification of the 670-990 cm$^{-1}$ region.

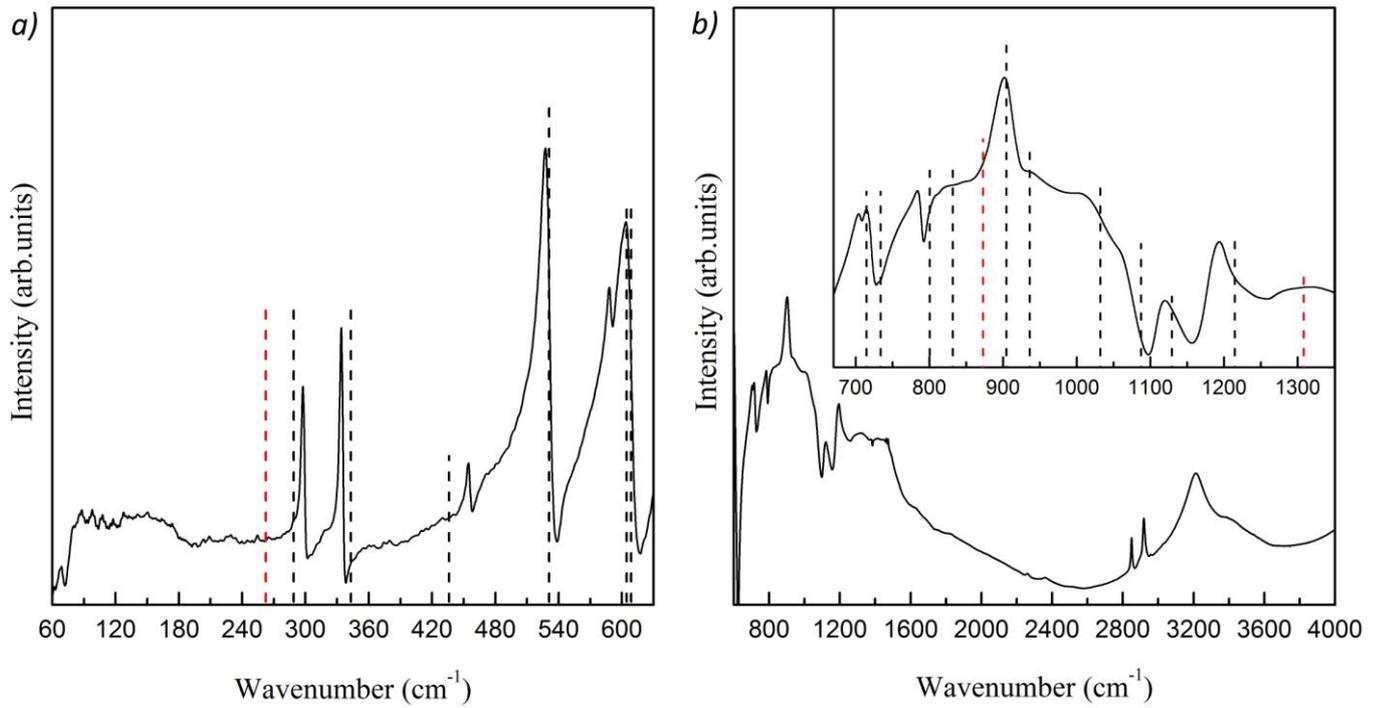

FIG. 7. Experimentally observed β-B$_2$O$_3$ phonons in far-infrared (*a*) and mid-infrared (*b*) regions at ambient conditions (Inset: magnification of the 640-1350 cm$^{-1}$ region). The phonon positions predicted by LCAO calculations are marked by the lines: the black dashed lines correspond to the experimentally observed phonons, the red dashed lines correspond to the non-experimentally observed phonons.

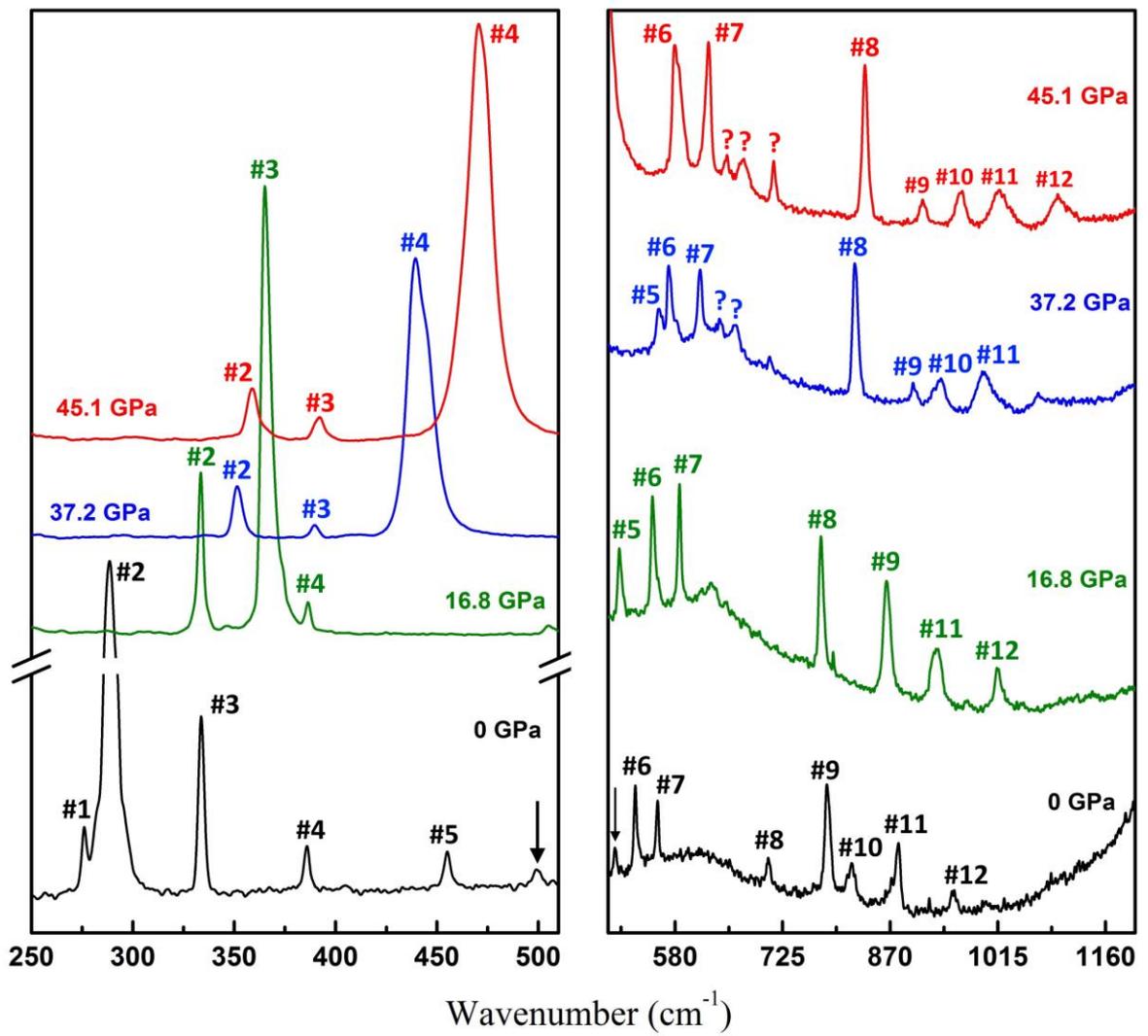

FIG. 8. Raman spectra of $\beta$-$B_2O_3$ at different pressures and room temperature. The Raman bands appeared during compression are marked by "#?", the Raman band related to a contamination is marked by arrow.

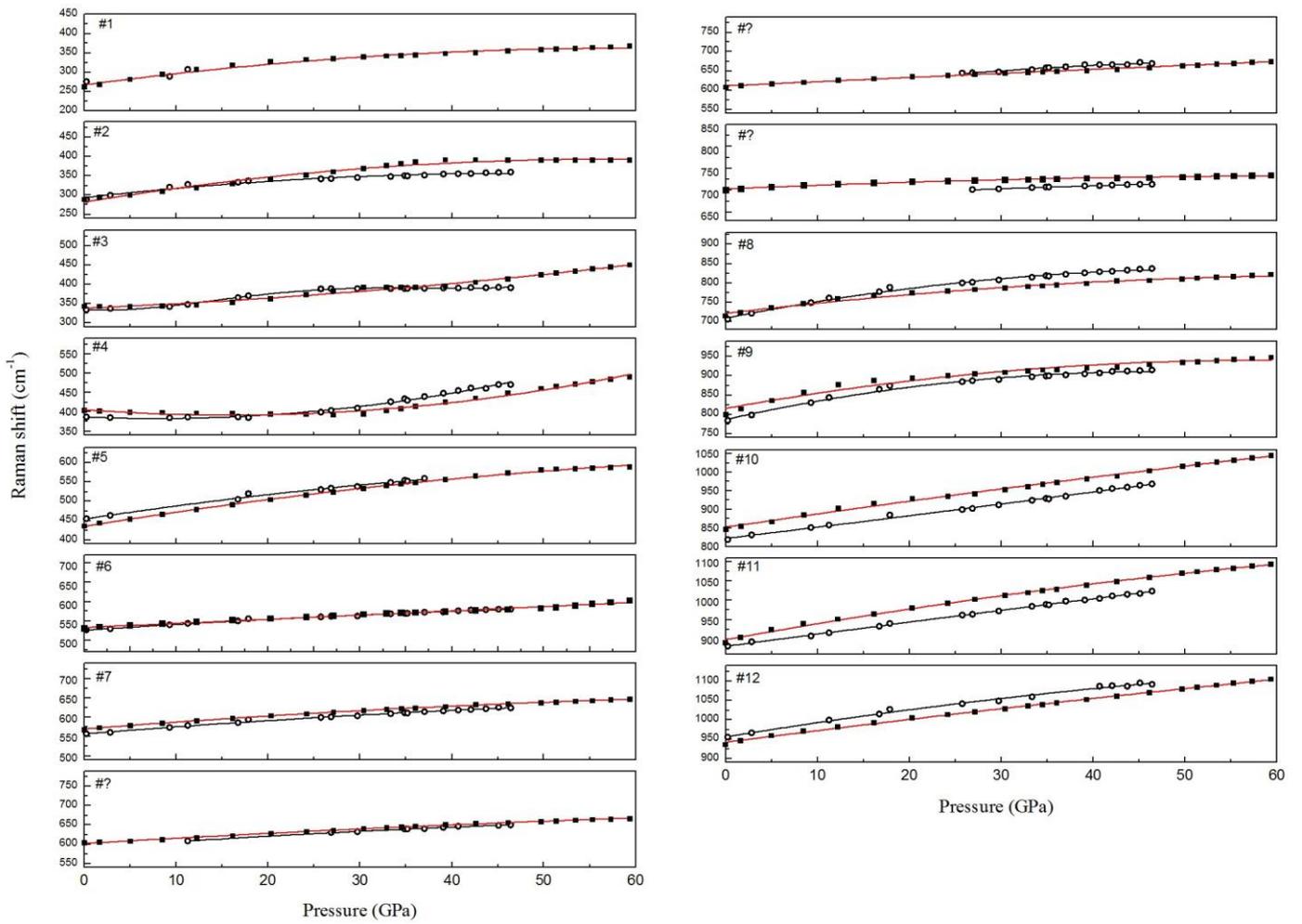

FIG. 9. Pressure dependencies of the experimentally observed (open circles) and theoretically calculated (solid squares) phonon mode frequencies. The black and red lines are quadratic least squares fits ($R^2 > 0.95$) to the experimental and theoretical data, respectively.